\documentclass[journal,10pt]{IEEEtran}

\IEEEoverridecommandlockouts

\usepackage[cmex10]{amsmath}
\usepackage[dvips]{graphicx}
\usepackage{cite}
\usepackage{amssymb}
\usepackage[eulergreek]{sansmath}
\usepackage{color}

\begin{document}

\title{Multiplexing Analysis of Millimeter-Wave Massive MIMO Systems}
\author{Dian-Wu Yue, Ha H. Nguyen and Shuai Xu
\thanks{Dian-Wu Yue is with the College of Information Science and
Technology, Dalian Maritime University, Dalian, Liaoning 116026,
China (e-mail: dwyue@dlmu.edu.cn), and also with the Department of Electrical and Computer Engineering, University of Saskatchewan, 57 Campus Drive, Saskatoon, SK, Canada S7N 5A9.}
\thanks{Ha H. Nguyen is with the Department of Electrical and Computer Engineering, University of Saskatchewan, 57 Campus Drive, Saskatoon, SK, Canada S7N 5A9 (e-mail: ha.nguyen@usask.ca).}
\thanks{Shuai Xu is with the College of Information Science and
Technology, Dalian Maritime University, Dalian, Liaoning 116026, China (e-mail: xu\_shuai@dlmu.edu.cn).}
}


\newcommand{\be}{\begin{equation}}
\newcommand{\ee}{\end{equation}}
\newcommand{\bee}{\begin{eqnarray}}
\newcommand{\eee}{\end{eqnarray}}
\newcommand{\nnb}{\nonumber}

\newcommand{\mo}{\mathbf{0}}
\newcommand{\mA}{\mathbf{A}}
\newcommand{\mB}{\mathbf{B}}
\newcommand{\mG}{\mathbf{G}}
\newcommand{\mH}{\mathbf{H}}
\newcommand{\mI}{\mathbf{I}}
\newcommand{\mR}{\mathbf{R}}
\newcommand{\mY}{\mathbf{Y}}
\newcommand{\mZ}{\mathbf{Z}}
\newcommand{\mD}{\mathbf{D}}
\newcommand{\mW}{\mathbf{W}}
\newcommand{\mF}{\mathbf{F}}
\newcommand{\mP}{\mathbf{P}}
\newcommand{\mU}{\mathbf{U}}
\newcommand{\mV}{\mathbf{V}}
\newcommand{\mSigma}{\mathbf{\Sigma}}

\newcommand{\my}{\mathbf{y}}
\newcommand{\mx}{\mathbf{x}}
\newcommand{\mz}{\mathbf{z}}

\newcommand{\mr}{\mathbf{r}}
\newcommand{\mt}{\mathbf{t}}
\newcommand{\mb}{\mathbf{b}}
\newcommand{\ma}{\mathbf{a}}

\newcommand{\mh}{\mathbf{h}}
\newcommand{\mw}{\mathbf{w}}
\newcommand{\mg}{\mathbf{g}}
\newcommand{\mf}{\mathbf{f}}
\newcommand{\mn}{\mathbf{n}}

\newcommand{\mv}{\mathbf{v}}
\newcommand{\ms}{\mathbf{s}}
\newcommand{\mmu}{\mathbf{u}}

\newcommand{\lf}{\left}
\newcommand{\ri}{\right}

\newtheorem{Lemma}{Lemma}
\newtheorem{Theorem}{Theorem}
\newtheorem{Corollary}{Corollary}
\newtheorem{Proposition}{Proposition}
\newtheorem{Example}{Example}
\newtheorem{Definition}{Definition}

\maketitle

\begin{abstract}
This paper is concerned with spatial multiplexing analysis for millimeter-wave (mmWave) massive MIMO systems. For a single-user mmWave system employing distributed antenna subarray architecture in which the transmitter and receiver consist of $K_t$ and $K_r$ subarrays, respectively, an asymptotic multiplexing gain formula is firstly derived when the numbers of antennas at subarrays go to infinity. Specifically, assuming that all subchannels have the same average number of propagation paths $\bar{L}$, the formula implies that by employing such a distributed antenna-subarray architecture, an exact average maximum multiplexing gain of $K_rK_t\bar{L}$ can be achieved. This result means that compared to the co-located antenna architecture, using the distributed antenna-subarray architecture can statistically scale up the maximum multiplexing gain proportionally to $K_rK_t$. In order to further reveal the relation between diversity gain and multiplexing gain, a simple characterization of the diversity-multiplexing tradeoff is also given. The multiplexing gain analysis is then extended to the multiuser scenario as well as the conventional partially-connected RF structure in the literature. Moreover, simulation results obtained with the hybrid analog/digital processing corroborate the analysis results.
\end{abstract}


\IEEEpeerreviewmaketitle


\section{Introduction}

Recently, millimeter-wave (mmWave) communication has gained considerable attention as a candidate technology for 5G mobile communication systems and beyond \cite{Rappaport13, Swindlehurst14, W.Roh14}. The main reason for this is the availability of vast spectrum in the mmWave band (typically 30-300 GHz) that is very attractive for high data rate communications. However, compared to communication systems operating at lower microwave frequencies (such as those currently used for 4G mobile communications), propagation loss in mmWave frequencies is much higher, in the orders-of-magnitude. Fortunately, given the much smaller carrier wavelengths, mmWave communication systems can make use of compact massive antenna arrays to compensate for the increased propagation loss.

Nevertheless, the large-scale antenna arrays together with high cost and large power consumption of the mixed analog/digital signal components makes it difficult to equip a separate radio-frequency (RF) chain for each antenna and perform all the signal processing in the baseband. Therefore, research on \emph{hybrid} analog-digital processing of precoder and combiner for mmWave communication systems has attracted very strong interests from both academia and industry \cite{Ayach12}$\;-$\cite{Molisch16}. In particular, a lot of work has been performed to address challenges in using a limited number of RF chains. For example, the authors in \cite{Ayach12} considered single-user precoding in mmWave massive MIMO systems and established the optimality of beam steering for both single-stream and multi-stream transmission scenarios. In \cite{Sohrabi16}, the authors showed that hybrid processing can realize any fully digital processing if the number of RF chains is twice the number of data streams. However, due to the fact that  mmWave signal propagation has an important feature of multipath sparsity in both the temporal and spatial domains \cite{Pi11, T.S.Rappaport13, Raghavan08, Raghavan11},  it is expected that the potentially available benefits of diversity and multiplexing are indeed not large if the deployment of the antenna arrays is co-located. In order to enlarge diversity/multiplexing gains in mmWave massive MIMO communication systems, this paper consider the use of a more general array architecture, called \emph{distributed antenna subarray architecture}, which includes lo-located array architecture as a special case. It is pointed out that distributed antenna systems have received strong interest as a promising technique to satisfy such growing demands for future wireless communication networks due to the increased spectral efficiency and expanded coverage \cite{Clark01}$\;-$\cite{Gimenez17}.

It is well known that diversity-multiplexing tradeoff (DMT) is a compact and convenient framework to compare different MIMO systems in terms of the two main and related system indicators: data rate and error performance \cite{Zheng03, Tse04, Zhao05, Narasimhan06, Yuksel07, Tse07}. This tradeoff was originally characterized by Zheng and Tse \cite{Zheng03} for MIMO communication systems operating over independent and identically distributed (i.i.d.) Rayleigh fading channels. The framework has then ignited a lot of interests in analyzing various communication systems and under different channel models. For a mmWave massive MIMO system, how to quantify the diversity and multiplexing performance and further characterize its DMT is a fundamental and open research problem. In particular, to the best of our knowledge, until now there is no unified multiplexing gain analysis for mmWave massive MIMO systems that is applicable to both co-located and distributed antenna array architectures.

To fill this gap, this paper investigates the multiplexing performance of mmWave massive MIMO systems with the proposed distributed subarray architecture. The focus is on the asymptotical multiplexing gain analysis in order to
find out the potential multiplexing advantage provided by multiple distributed antenna arrays. The obtained analysis can be used conveniently to compare various mmWave massive MIMO systems with different distributed antenna array structures.

The main contributions of this paper are summarized as follows:
\begin{itemize}
  \item{For a single-user system with the proposed distributed subarray architecture, a multiplexing gain expression is obtained when the number of antennas at each subarray increases without bound. This expression clearly indicates that one can obtain a large multiplexing gain by employing the distributed subarray architecture. }
  \item{A simple DMT characterization is further given. It can reveal the relation between diversity gain and multiplexing gain and let us obtain insights to understand the overall resources provided by the distributed antenna architecture.}
  \item {The multiplexing gain analysis is then extended to the multiuser scenario with downlink and uplink transmission, as well as the single-user system employing the conventional partially-connected RF structure based distributed subarrays.}
  \item {Simulation results are provided to corroborate the analysis results and show that the distributed subarray architecture yields significantly better multiplexing performance than the co-located single-array architecture.}
\end{itemize}

The remainder of this paper is organized as follows. Section II describes the massive MIMO system model and hybrid processing with the distributed subarray architecture in mmWave fading channels. Section III and Section IV provides the asymptotical achievable rate analysis and the multiplexing gain analysis for the single-user mmWave system, respectively. In Section V and VI, the multiplexing gain analysis is extended to the scenario with the partially-connected RF architecture and the multiuser scenario, respectively. Section VII concludes the paper.

Throughout this paper, the following notations are used. Boldface upper and lower case letters denote matrices and column vectors, respectively. The superscripts $(\cdot)^T$  and  $(\cdot)^H$ stand for transpose and conjugate-transpose, respectively. $\mathrm{diag}\{a_1,a_2,\ldots,a_N\}$ stands for a diagonal matrix with diagonal elements $\{a_1,a_2,\ldots,a_N\}$. The expectation operator is denoted by $\mathbb{E}(\cdot)$.
$[\mA]_{ij}$ gives the $(i,j)$th  entry of matrix $\mA$. $\mA \bigotimes\mB$ is the Kronecker product of $\mA$ and $\mB$. We write a function $a(x)$ of $x$ as $o(x)$ if $\lim_{x \to 0}a(x)/x=0$. We use $(x)^+$ to denote $\max \{0, x\}$. Finally, $\mathcal{CN}(0, 1)$  denotes a circularly symmetric complex Gaussian random variable with zero mean and unit variance.

\section{System Model}

\begin{figure*}[t]
\centering
\includegraphics[scale = 0.8]{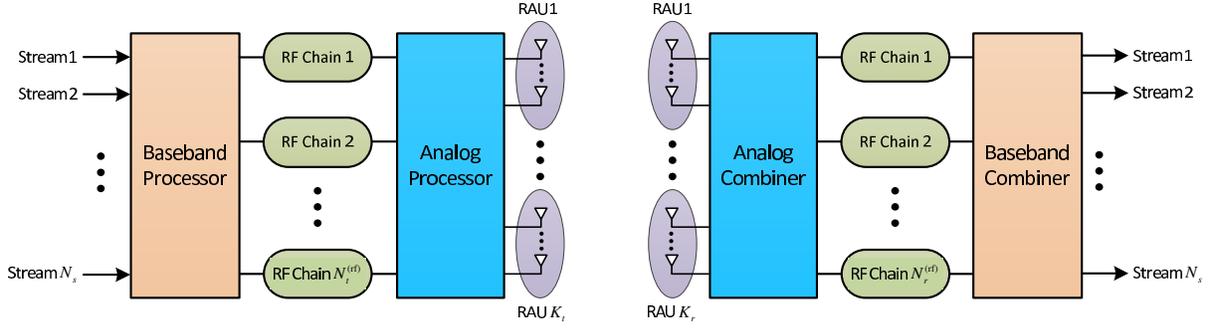}
\caption{Block diagram of a mmWave massive MIMO system with distributed antenna arrays.}
\label{SYS1}
\end{figure*}

\begin{figure}[t]
\centering
\includegraphics[scale = 0.6]{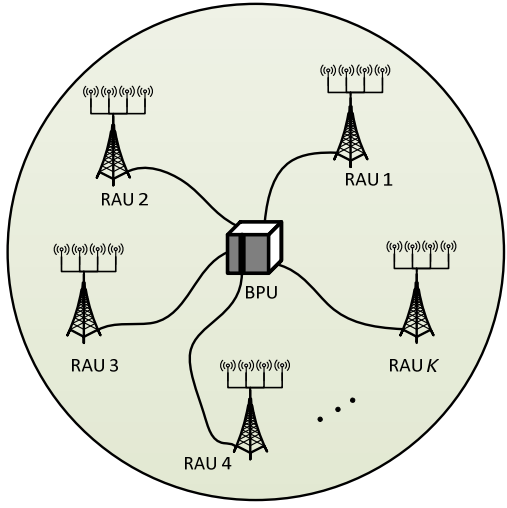}
\caption{Illustration of distributed antenna array deployment.}
\label{SYS2}
\end{figure}

Consider a single-user mmWave massive MIMO system as shown in Fig. \ref{SYS1}. The transmitter is equipped with a distributed antenna array to send $N_s$ data streams to a receiver, which is also equipped with a distributed antenna array. Here, a distributed antenna array means an array consisting of several remote antenna units (RAUs) (i.e., antenna subarrays) that are distributively located, as depicted in Fig. \ref{SYS2}. Specifically, the antenna array at the transmitter consists of $K_t$ RAUs, each of which has $N_t$ antennas and is connected to a baseband processing unit (BPU) by fiber. Likewise, the distributed antenna array at the receiver consists of $K_r$ RAUs, each having $N_r$ antennas and also being connected to a BPU by fibers. Such a MIMO system shall be referred to as a $(K_t, N_t, K_r, N_r)$ distributed MIMO (D-MIMO) system. When $K_t=K_r=1$, the system reduces to a conventional co-located MIMO (C-MIMO) system.

The transmitter accepts as its input $N_s$ data streams and is equipped with $N_t^{({\rm rf})}$ RF chains, where $N_s \leq N_t^{({\rm rf})} \leq N_t K_t $. Given $N_t^{({\rm rf})}$ transmit RF chains, the transmitter can apply a low-dimension  $N_t^{({\rm rf})} \times N_s$ baseband precoder, $\mW_t$, followed by a high-dimension  $K_tN_t \times N_t^{({\rm rf})}$ RF precoder, $\mF_t$. Note that amplitude and phase modifications are feasible for the baseband precoder $\mW_t$, while only phase changes can be made by the RF precoder $\mF_t$ through the use of variable phase shifters and combiners. The transmitted signal vector can be written as
\be \mx = \mF_t\mW_t\mP_t^{1/2}\ms, \ee
where $\ms$ is the $N_s  \times  1$ symbol vector such that $\mathbb{E}[\ms\ms^H] = \mI_{N_s}$, and $\mP_t=[p_{ij}]$ is a diagonal power allocation matrix with $\sum_{l=1}^{N_s}p_{ll}=P$. Thus $P$ represents the average total input power. Considering a narrowband block fading channel, the $K_rN_r \times 1$ received signal vector is
\be \label{my} \my =\mH\mF_t\mW_t\mP_t^{1/2}\ms + \mn  \ee
where $\mH$ is $K_rN_r \times K_tN_t$ channel matrix and $\mn$ is a $K_rN_r \times 1$ vector consisting of i.i.d. $\mathcal{CN}(0, 1)$ noise samples. Throughout this paper, $\mH$ is assumed known to both the transmitter and receiver. Given that $N_r^{({\rm rf})}$ RF chains (where $N_s \leq N_r^{({\rm rf})} \leq N_rK_r$) are used at the receiver to detect the $N_s$ data streams, the processed signal is given by
\be  \label{mz} \mz =\mW_r^H\mF_r^H\mH\mF_t\mW_t\mP_t^{1/2}\ms + \mW_r^H\mF_r^H\mn  \ee
where $\mF_r$ is the $ K_rN_r \times N_r^{({\rm rf})} $ RF combining matrix, and $\mW_r$ is the $N_r^{({\rm rf})} \times N_s $ baseband combining matrix. When Gaussian symbols are transmitted over the mmWave channel, the
the system achievable rate is expressed as
\be R=\log_2|\mI_{N_s}+\mR_n^{-1}\mW_r^H\mF_r^H\mH\mF_t\mW_t\mP_t\mW_t^H\mF_t^H\mH^H\mF_r\mW_r|\ee
where $\mR_n=\mW_r^H\mF_r^H\mF_r\mW_r$.

Furthermore, according to the architecture of RAUs at the transmitting and receiving ends, $\mH$ can be written as
\be \label{mH}
    \mH=\left[\begin{array}{lll}
   \sqrt{g_{11}}\mH_{11} & \cdots & \;\;\sqrt{g_{1K_t}}\mH_{1K_t} \\
    \;\;\;\;\vdots & \ddots & \;\;\;\;\;\;\;\;\vdots \\
  \sqrt{g_{K_r1}}\mH_{K_r1} & \cdots & \sqrt{g_{K_rK_t}}\mH_{K_rK_t}
  \end{array}
   \right].
 \ee
In the above expression,  $g_{ij}$ represents the large scale fading effect between the $i$th RAU at the receiver and the $j$th RAU at the transmitter, which is assumed to be constant over many coherence-time intervals. The normalized subchannel matrix $\mH_{ij}$ represents the MIMO channel between the $j$th RAU at the transmitter and the $i$th RAU at the receiver. We assume that all of $\{\mH_{ij}\}$ are independent mutually each other.

A clustered channel model based on the extended Saleh-Valenzuela model is often used in mmWave channel modeling and standardization \cite{Ayach12} and it is also adopted in this paper. For simplicity of exposition, each scattering cluster is assumed to contribute a single propagation path.\footnote{This assumption can be relaxed to account for clusters with finite angular spreads and the results obtained in this paper can be readily extended for such a case.} Using this model, the subchannel matrix $\mH_{ij}$ is given by
\be \mH_{ij}=\sqrt{\frac{N_tN_r}{L_{ij}}}\sum_{l=1}^{L_{ij}}\alpha_{ij}^l\ma_r(\phi^{rl}_{ij},\theta^{rl}_{ij})\ma_t(\phi^{tl}_{ij},\theta^{tl}_{ij})^H,   \ee
where $L_{ij}$ is the number of propagation paths, $\alpha_{ij}^l$ is the complex gain of the $l$th ray, and $\phi^{rl}_{ij}$ ($\theta^{rl}_{ij}$) and $\phi^{tl}_{ij}$ ($\theta^{tl}_{ij}$) are its random azimuth (elevation) angles of arrival and departure, respectively. Without loss of generality, the complex
gains $\alpha_{ij}^l$ are assumed to be $\mathcal{CN}(0, 1)$. \footnote{The different variances of $\alpha_{ij}^l$ can easily accounted for by absorbing into the large scale fading coefficients $g_{ij}$.} The vectors $\ma_r(\phi^{rl}_{ij},\theta^{rl}_{ij})$ and $\ma_t(\phi^{tl}_{ij},\theta^{tl}_{ij})$ are the normalized receive/transmit array response vectors at the corresponding angles of arrival/departure. For an $N$-element uniform linear array (ULA) , the
array response vector is \be \ma^{\mathrm{ULA}}(\phi)=\frac{1}{\sqrt{N}}\left[1,{\mathrm e}^{j2\pi\frac{d_u}{\lambda}\sin(\phi)},\ldots,{\mathrm e}^{j2\pi(N-1)\frac{d_u}{\lambda}\sin(\phi)}\right]^T \ee
where $ \lambda$ is the wavelength of the carrier and $d_u$ is the inter-element spacing. It is pointed out that the angle
$\theta$ is not included in the argument of $\ma^{\mathrm{ULA}}$ since the response for an ULA is independent of the elevation angle. In contrast, for a uniform planar array (UPA),  which is composed of $N_h$ and $N_v$ antenna elements in the horizontal and vertical directions, respectively, the array response vector is represented by
\be  \ma^{\mathrm{UPA}}(\phi, \theta)=\ma^{\mathrm{ULA}}_h(\phi)\otimes \ma^{\mathrm{ULA}}_v(\theta), \ee
where
\be \ma^{\mathrm{ULA}}_h(\phi)=\frac{1}{\sqrt{N_h}}\left[1,{\mathrm e}^{j2\pi\frac{d_h}{\lambda}\sin(\phi)},\ldots,{\mathrm e}^{j2\pi(N_h-1)\frac{d_h}{\lambda}\sin(\phi)}\right]^T \ee
and
\be \ma^{\mathrm{ULA}}_v(\theta)=\frac{1}{\sqrt{N_v}}\left[1,{\mathrm e}^{j2\pi\frac{d_v}{\lambda}\sin(\theta)},\ldots,{\mathrm e}^{j2\pi(N^v-1)\frac{d_v}{\lambda}\sin(\theta)}\right]^T. \ee

\section{Asymptotic Achievable Rate Analysis}

From the structure and definition of the channel matrix $\mH$ in Section II, there is a total of $L_s=\sum_{i=1}^{K_r}\sum_{j=1}^{K_t}L_{ij}$ propagation paths. Naturally, $\mH$ can be decomposed into a sum of $L_s$ rank-one matrices, each corresponding to one propagation path. Specifically, $\mH$ can be rewritten as
\be \mH=\sum_{i=1}^{K_r}\sum_{j=1}^{K_t}\sum_{l=1}^{L_{ij}}\tilde{\alpha}_{ij}^l\tilde{\ma}_r(\phi^{rl}_{ij},\theta^{rl}_{ij})\tilde{\ma}_t^H(\phi^{tl}_{ij},\theta^{tl}_{ij}),   \ee
where \be  \tilde{\alpha}_{ij}^l =\sqrt{g_{ij}\frac{N_tN_r}{L_{ij}}}\alpha_{ij}^l, \ee  $\tilde{\ma}_r(\phi^{rl}_{ij},\theta^{rl}_{ij})$ is a $K_rN_r \times 1$ vector whose $b$th entry is defined as
\be \label{rrr}[\tilde{\ma}_r(\phi^{rl}_{ij},\theta^{rl}_{ij})]_b=\left\{\begin{array}{ll}
[\ma_r(\phi^{rl}_{ij},\theta^{rl}_{ij})]_{b-(i-1)N_r}, & b\in Q_i^r\\
0, &  b \notin Q_i^r
     \end{array}
     \right. \ee
where $Q_i^r=((i-1)N_r, iN_r]$. And $\tilde{\ma}_t(\phi^{tl}_{ij},\theta^{tl}_{ij})$ is a $K_tN_t \times 1$ vector whose $b$th entry is defined as
\be \label{ttt} [\tilde{\ma}_t(\phi^{tl}_{ij},\theta^{tl}_{ij})]_b=\left\{\begin{array}{ll}
[\ma_t(\phi^{tl}_{ij},\theta^{tl}_{ij})]_{b-(j-1)N_t}, &  b\in Q_j^t\\
0, & b\notin Q_j^t.
     \end{array}
     \right. \ee
where $Q_j^t=((j-1)N_t, jN_t]$. Regarding $\{\tilde{\ma}_r(\phi^{rl}_{ij},\theta^{rl}_{ij})\}$ and $\{\tilde{\ma}_t(\phi^{tl}_{ij},\theta^{tl}_{ij})\}$, we have the following lemma from \cite{Yue17}.

\begin{Lemma} Suppose that the antenna configurations at all RAUs are either ULA or UPA. Then all $L_s$ vectors $\{\tilde{\ma}_r(\phi^{rl}_{ij},\theta^{rl}_{ij})\}$ are orthogonal to each other when $N_r \to \infty$. Likewise, all $L_s$ vectors $\{\tilde{\ma}_t(\phi^{tl}_{ij},\theta^{tl}_{ij})\}$ are orthogonal to each other when $N_t \to \infty$.
\end{Lemma}

Mathematically, the distributed massive MIMO system can be considered as a co-located massive MIMO system with $L_s$ paths that have complex gains $\{\tilde{\alpha}_{ij}^l\}$, receive array response vectors $\{\tilde{\ma}_r(\phi^{rl}_{ij},\theta^{rl}_{ij})\}$ and transmit response vectors $\{\tilde{\ma}_t(\phi^{tl}_{ij},\theta^{tl}_{ij})\}$. Furthermore, order all paths in a decreasing order of the absolute values of the complex gains $\{\tilde{\alpha}_{ij}^l\}$. Then the channel matrix can be written as
\be  \mH=\sum_{l=1}^{L_s}\tilde{\alpha}^l\tilde{\ma}_r(\phi^{rl},\theta^{rl})\tilde{\ma}_t(\phi^{tl},\theta^{tl})^H,  \ee
where $\tilde{\alpha}^1\geq \tilde{\alpha}^2\geq \cdots \geq \tilde{\alpha}^{L_s}$.

One can rewrite $\mH$ in a matrix form as
\be  \mH=\mA_r\mD\mA_t^H  \ee
where $\mD$ is a $L_s \times L_s$ diagonal matrix with $[\mD]_{ll}=\tilde{\alpha}^l$, and $\mA_r$ and  $\mA_t$ are defined as follows:
\be  \mA_r=[\tilde{\ma}_r(\phi^{r1},\theta^{r1}),\ldots,\tilde{\ma}_r(\phi^{rL_s},\theta^{rL_s})]  \ee
and
      \be  \mA_t=[\tilde{\ma}_t(\phi^{t1},\theta^{t1}),\ldots,\tilde{\ma}_t(\phi^{tL_s},\theta^{tL_s})].  \ee
Since both $\{\tilde{\ma}_r(\phi^{rl},\theta^{rl})\}$ and $\{\tilde{\ma}_t(\phi^{tl},\theta^{tl})\}$ are orthogonal vector sets when $N_r \to \infty$ and $N_t \to \infty$, $\mA_r$ and  $\mA_t$  are asymptotically unitary matrices. Then one can form a singular value decomposition (SVD) of matrix $\mH$ as
\be \label{SVD}\mH=\mU\mSigma\mV^H=[\mA_r|\mA_r^{\bot}]\mSigma [\tilde{\mA}_t|\tilde{\mA}_t^{\bot}]^H  \ee
where $\mSigma$ is a diagonal matrix containing all singular values on its diagonal, i.e.,
 \be [\mSigma]_{ll}=\left\{\begin{array}{ll}
 |\tilde{\alpha}^l|, & \mbox{for}\; 1 \leq  l\leq L_s\\
0, & \mbox{for}\; l>L_s
     \end{array}
     \right. \ee
and the submatrix $\tilde{\mA}_t$ is defined as
\be  \tilde{\mA}_t=[{\mathrm e}^{j\psi_1}\tilde{\ma}_t(\phi^{t1},\theta^{t1}),\ldots,{\mathrm e}^{j\psi_{L_s}}\tilde{\ma}_t(\phi^{tL_s},\theta^{tL_s})]   \ee
where $\psi_l$ is the phase of complex gain $\tilde{\alpha}^l$ corresponding to the $l$th path. Based on (\ref{SVD}), the optimal precoder and combiner are chosen, respectively, as
\be  \label{Ft} [\mF_t\mW_t]_\mathrm{opt}=[{\mathrm e}^{j\psi_1}\tilde{\ma}_t(\phi^{t1},\theta^{t1}),\ldots,{\mathrm e}^{j\psi_{L_s}}\tilde{\ma}_t(\phi^{tN_s},\theta^{tN_s})]  \ee
and
\be  \label{Fr}[\mF_r\mW_r]_\mathrm{opt}=[\tilde{\ma}_r(\phi^{r1},\theta^{r1}),\ldots,\tilde{\ma}_r(\phi^{rN_s},\theta^{rN_s})].  \ee

To summarize, when $N_t$ and  $N_r$ are large enough, the massive MIMO system can employ the optimal precoder and combiner given in (\ref{Ft}) and (\ref{Fr}), respectively.

For a given $l$, there are $l^\prime$, $i^\prime$ and $j^\prime$ such that $\tilde{\alpha}^l=\tilde{\alpha}_{i^\prime j^\prime}^{l^\prime} =\sqrt{g_{i^\prime j^\prime}\frac{N_tN_r}{L_{i^\prime j^\prime}}}\alpha_{i^\prime j^\prime}^{l^\prime}$. So we introduce two notations: \be \tilde{\gamma}_l=p_{ll}g_{i^\prime j^\prime}\frac{N_tN_r}{L_{i^\prime j^\prime}}\ee and \be\tilde{\beta}_l=\alpha_{i^\prime j^\prime}^{l^\prime}. \ee
Then it follows from the above SVD analysis that the instantaneous SNR of the $l$th data stream is given by
\be  \mathrm{SNR}_l=p_{ll}|\tilde{\alpha}^l|^2=\tilde{\gamma}_l|\tilde{\beta}_l|^2, \;\; l=1,2,\ldots, N_s.        \ee
So we obtain another lemma.
\begin{Lemma} Suppose that both sets $\{\tilde{\ma}_r(\phi^{rl}_{ij},\theta^{rl}_{ij})\}$ and $\{\tilde{\ma}_t(\phi^{tl}_{ij},\theta^{tl}_{ij})\}$ are orthogonal vector sets when $N_r \to \infty$ and $N_t \to \infty$. Let $N_s \leq L_s$.  In the limit of large $N_t$ and $N_r$,  then the system achievable rate is given by
\be \label{R} R=\sum_{l=1}^{Ns}\log_2(1+\tilde{\gamma}_l|\tilde{\beta}_l|^2).     \ee
\end{Lemma}

{\em Remark 1:}  (\ref{Ft}) and (\ref{Fr}) indicate that when $N_t$ and $N_r$ is large enough, the
optimal precoder and combiner can be implemented fully in RF using phase shifters \cite{Ayach12}. Furthermore, (\ref{rrr}) and (\ref{ttt}) imply that for
each data stream only a couple of RAUs needs the operation of phase shifters at each channel
realization.

{\em Remark 2:}  By using the optimal power allocation (i.e., the well-known waterfilling power allocation \cite{Telatar99}), the system can achieve a maximum achievable rate, which is denoted as $R_o$.  We use $R_e(P/N_s)$  to denote the achievable rate obtained by using the equal power allocation, namely, $p_{ll}=\frac{P}{N_s}, \; l=1,2,\ldots, N_s.$ Then
\be  \label{RRR} R_e(P/N_s) \leq R_o \leq R_e((PN_s)/N_s)=R_e(P).\ee
By doing expectation operation on (\ref{RRR}),   (\ref{RRR}) becomes,
\be \label{Rw} \bar{R}_e(P/N_s) \leq \bar{R}_o \leq \bar{R}_e(P).\ee

In what follows, we derive an asymptotic expression of the ergodic achievable rate with the equal power allocation, $\bar{R}_e(P/N_s)$ (or $\bar{R}_e$ for simplicity).  For this reason, we need to define an integral function
\bee \Delta(x) & =&\int_0^{+\infty}\log_2(1+t)e^{-t/x}\frac{1}{x}dt \nnb \\
                &=& \log_2(e)e^{1/x}E_1(1/x) \eee
where $E_1(\cdot)$ is the exponential integral of a first-order function defined as \cite{Simon05}, \cite{Gradshteyn94} \bee E_1(y)&=&\int_1^{+\infty}\frac{e^{-yt}}{t}dt\nnb\\
&=& -E+\ln(y)-\sum_{k=1}^{\infty}\frac{(-y)^k}{k \cdot k!}\eee
with $E$ being the Euler constant.

\begin{Theorem} Suppose that both sets $\{\tilde{\ma}_r(\phi^{rl}_{ij},\theta^{rl}_{ij})\}$ and $\{\tilde{\ma}_t(\phi^{tl}_{ij},\theta^{tl}_{ij})\}$ are orthogonnal vector sets when $N_r \to \infty$ and $N_t \to \infty$. Let $N_s \leq L_s$ and  $\tilde{\gamma}_1=\tilde{\gamma}_2=\ldots=\tilde{\gamma}_{L_s}=\tilde{\gamma}$.  Then in the limit of large $N_t$ and $N_r$, the ergodic achievable rate with homogeneous coefficient set $ \{\tilde{\gamma}_l\} $, denoted $\bar{R}_{eh}$, is given by
\be  \label{Re}\bar{R}_{eh}=\sum_{l=1}^{N_s} \sum_{k=0}^{L_s-l}\frac{(-1)^{L_s-l-k}L_s!}{(L_s-l)!(l-1)!}\binom{L_s-l}{k}\frac{\Delta(\frac{\tilde{\gamma}}{L_s-k})}{L_s-k} .    \ee
When $N_s = L_s$, $\bar{R}_{eh}$ can be simplified to
\be  \label{Ree} \bar{R}_{eh}=L_s \Delta(\tilde{\gamma}). \ee
\end{Theorem}
{\em Proof:} Due to the assumptions  that each complex gain $\alpha_{ij}^l$ is $\mathcal{CN}(0, 1)$ and the coefficient set $ \{\tilde{\gamma}_l\} $ is homogeneous, thus the instantaneous SNRs in the $L_s$ available data streams are i.i.d.. Let $F(\gamma)$ and $f(\gamma)$ denote the cumulative distribution function (CDF) and the probability density function (PDF) of the unordered instantaneous SNRs, respectively. Then $\tilde{\gamma}$ is just the average receive SNR of each data stream. Furthermore,  $F(\gamma)$ and $f(\gamma)$ can be written as
\be \label{Ff} F(\gamma)=1-{\mathrm e}^{-\frac{\gamma}{\tilde{\gamma}}},\;\; f(\gamma)=\frac{1}{\tilde{\gamma}}{\mathrm e}^{-\frac{\gamma}{\tilde{\gamma}}}. \ee
For the $l$th best data stream, based on the theory of order statistics \cite{David81}, the PDF of the instantaneous receive SNR at the receiver, denoted $\gamma_l$, is given by
\be \label{fff} f_{l:L_s}(\gamma_l)=\frac{L_s!}{(L_s-l)!(l-1)!} [F(\gamma_l)]^{L_s-l}[1-F(\gamma_l)]^{l-1}f(\gamma_l).\ee
Inserting (\ref{Ff}) into (\ref{fff}), we have that
\be f_{l:L_s}(\gamma_l)=\sum_{k=0}^{L_s-l}\frac{L_s!(-1)^{L_s-l-k}}{(L_s-l)!(l-1)!} \binom{L_s-l}{k}\frac{e^{-\gamma_l (L_s-k)/\tilde{\gamma}}}{\tilde{\gamma}}.\ee
By the definition of the function $\Delta(\cdot)$,  the ergodic available rate for the $l$th data stream can therefore be written as
\bee &&R_{eh}^{(l)}= \int_0^{+\infty}\log_2(1+\gamma_l)f_{l:L_s}(\gamma_l)d\gamma_l \nnb \\
&&= \sum_{k=0}^{L_s-l}\binom{L_s-l}{k}\frac{L_s!(-1)^{L_s-l-k}}{(L_s-l)!(l-1)!} g_k(\tilde{\gamma} )\nnb \\
 &&=\sum_{k=0}^{L_s-l}\binom{L_s-l}{k} \frac{L_s!(-1)^{L_s-l-k}}{(L_s-l)!(l-1)!} \frac{\Delta(\frac{\tilde{\gamma}}{L_s-k})}{L_s-k}\eee
 where \be g_k(\tilde{\gamma} )= \int_0^{+\infty}\log_2(1+\gamma_l)\frac{e^{-\gamma_l(L_s-k)/\tilde{\gamma} }}{\tilde{\gamma} }d\gamma_l.\ee
 So we can obtain the desired result (\ref{Re}).

  Finally, when $N_s = L_s$, we can readily prove (\ref{Ree}) with the help of the knowledge of unordered statistics.  \hfill $\square$

{\em Remark 3:} Now let $N_s = L_s$ and assume that $L_{ij}=L$ for any $i$ and $j$ (i.e., all subchannels $\mH_{ij}$ have the same number of propagation paths). When $N_r \to \infty$ and $N_t \to \infty$, the ergodic achievable rate of the distributed MIMO system,  $\bar{R}_{eh}$,  can be rewritten as
 \be \bar{R}_{eh}(K_t, K_r)=K_t K_rL \Delta(\tilde{\gamma}(K_t, K_r)).  \ee
Furthermore, consider a co-located MIMO system in which the numbers of transmit and receiver antennas are equal to $K_tN_t$ and $K_rN_r$, respectively. Assume that the number of propagation paths is equal to $L$. Then its asymptotic ergodic achievable rate can be expressed as
\be  \bar{R}_{eh}(1, 1)=L \Delta(K_t K_r\tilde{\gamma}(K_t, K_r)). \ee

{\em Remark 4:} Generally, the coefficient set $ \{\tilde{\gamma}_l\} $ is inhomogeneous. Let $ \tilde{\gamma}_{max}=\max\{\tilde{\gamma}_l\} $ and $ \tilde{\gamma}_{min}=\min\{\tilde{\gamma}_l\} $. Then
the ergodic achievable rate with inhomogeneous coefficient set $ \{\tilde{\gamma}_l\} $, $\bar{R}_{e}$, has the following upper and lower bounds:
\be \label{Rh} \bar{R}_{eh}(\tilde{\gamma}_{min})\leq \bar{R}_{e}(\{\tilde{\gamma}_l\})\leq \bar{R}_{eh}(\tilde{\gamma}_{max}). \ee

 Assuming that $N_t^{({\rm rf})}=N_r^{({\rm rf})}=2N_s$ \cite{Sohrabi16} and $N_r=N_t=N$, Fig.3 plots the ergodic achievable rate curves versus $\mbox{SNR}_a=\tilde{\gamma}$ for different numbers of antennas, $N=5, 10, 50$. In Fig.3, we set $N_s=6$, $K_r=K_t=2$, and $L_{ij}=L=3$. As expected, it can be seen that the rate performance is improved as $N$ increases. Obviously, the rate curve with $N=10$ is very close to the curve with limit results obtained based on (30) while the rate curve with $N=50$ is almost the same as the curve with limit results. This verifies Theorem 1.

\begin{figure}[t]
\centering
\includegraphics[scale = 0.6]{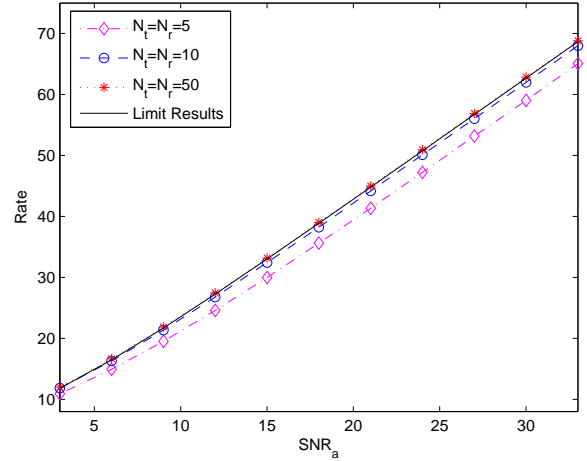}
\caption{Rate versus $\mbox{SNR}_a$ for different numbers of antennas.}
\label{SIM3}
\end{figure}

\section{Multiplexing Gain Analysis and Diversity-Multiplexing Tradeoff}

\subsection{Multiplexing Gain Analysis}
\begin{Definition} Let $\bar{\gamma}=\frac{1}{L_s}\sum_{l=1}^{Ls}\tilde{\gamma}_l$. The distributed MIMO system is said to achieve spatial multiplexing gain $G_m$ if its ergodic date rate with optimal power allocation satisfies
\be G_m(\bar{R}_o)=\lim_{\bar{\gamma} \to \infty}\frac{\bar{R}_o(\bar{\gamma})}{\log_2\bar{\gamma}}.\ee
\end{Definition}

\begin{Theorem} Assume that both sets $\{\tilde{\ma}_r(\phi^{rl}_{ij},\theta^{rl}_{ij})\}$ and $\{\tilde{\ma}_t(\phi^{tl}_{ij},\theta^{tl}_{ij})\}$ are orthogonal vector sets when $N_r $ and $N_t$ are very large. Assume that $N_r $ and $N_t$ are always very large but fixed and finite when $\bar{\gamma} \to \infty$. Let $N_s \leq L_s$. Then the spatial multiplexing gain is given by
\be G_m(\bar{R}_o)=N_s. \ee
\end{Theorem}

{\em Proof:} We first consider the simple homogeneous case with  $\tilde{\gamma}_1=\tilde{\gamma}_2=\ldots=\tilde{\gamma}_{L_s}=\tilde{\gamma}$ and derive the spatial multiplexing gain with respect to $\bar{R}_{eh}$.
In this case, $\bar{\gamma}=\tilde{\gamma}$. Obviously,
\be G_m(\bar{R}_{eh})=\lim_{\bar{\gamma} \to \infty}\frac{\bar{R}_{eh}(\bar{\gamma})}{\log_2\bar{\gamma}}=\sum_{l=1}^{N_s}\lim_{\bar{\gamma} \to \infty}\frac{\bar{R}_{eh}^{l}(\bar{\gamma})}{\log_2\bar{\gamma}}. \ee
For the $l$th data stream, under the condition of very large $N_t$ and $N_r$, the individual ergodic rate can be written as
\be R_{eh}^{(l)}(\bar{\gamma})=\mathbb{E}\log_2(1+\bar{\gamma}|\tilde{\beta}_l|^2).  \ee
Noting that \be \mathbb{E}\log_2(|\tilde{\beta}_l|^2)\leq \mathbb{E}\log_2(1+|\tilde{\beta}_l|^2)=R_{eh}^{(l)}(1)\ee and $R_{eh}^{(l)}(1)$ is a finite value, we can have that
\bee   G_m^{(l)}(\bar{R}_{eh})&=&\lim_{\bar{\gamma} \to \infty}\frac{R_{eh}^{(l)}(\bar{\gamma})}{\log_2\bar{\gamma}}\nnb\\
&=& \lim_{\bar{\gamma} \to \infty}\frac{\log_2\bar{\gamma}+\mathbb{E}\log_2(|\tilde{\beta}_l|^2)}{\log_2\bar{\gamma}} \nnb\\
&=& 1. \eee
Therefore, $G_m(\bar{R}_{eh})=\sum_{l=1}^{N_s}G_m^{(l)}(\bar{R}_{eh})=N_s$.

Now we consider the general inhomogeneous case with equal power allocation. Because $ c_{min}=\tilde{\gamma}_{min}/\bar{\gamma}$ and $ c_{max}=\tilde{\gamma}_{max}/\bar{\gamma}$ be finite when $\bar{\gamma} \to \infty$.
Consequently, it readily follows that both of the two systems with the achievable rates  $\bar{R}_{eh}(\tilde{\gamma}_{min})$ and $\bar{R}_{eh}(\tilde{\gamma}_{max})$ can achieve a multiplexing gain of $G_m(\bar{R}_{eh})=N_s$. So we conclude from (\ref{Rh}) that the distributed MIMO system with the achievable rate $\bar{R}_{e}(\bar{\gamma})$ can achieve a multiplexing gain of $G_m(\bar{R}_{e})=N_s$.

Finally, it can be readily shown that the system with the optimal achievable rate $\bar{R}_{o}(\bar{\gamma})$ can only achieve multiplexing gain $G_m(\bar{R}_{o})=N_s$ since both of the equal power allocation systems with the achievable rates  $\bar{R}_{e}(P/N_s)$ and $\bar{R}_{e}(P)$ have the same spatial multiplexing gain $N_s$. \hfill $\square$

\begin{Corollary} Assume that for any $i$ and $j$, the average number of propagation paths $\bar{L}_{ij}=\bar{L}$ . Then the distributed massive MIMO system can obtain an average maximum spatial multiplexing gain of $K_rK_t \bar{L}$.
\end{Corollary}

{\em Remark 5:}  Corollary 1 means that compared to the co-located antenna architecture, using the distributed antenna-subarray architecture can statistically scale up the maximum multiplexing gain proportionally to $K_rK_t$.

\begin{figure}[t]
\centering
\includegraphics[scale = 0.6]{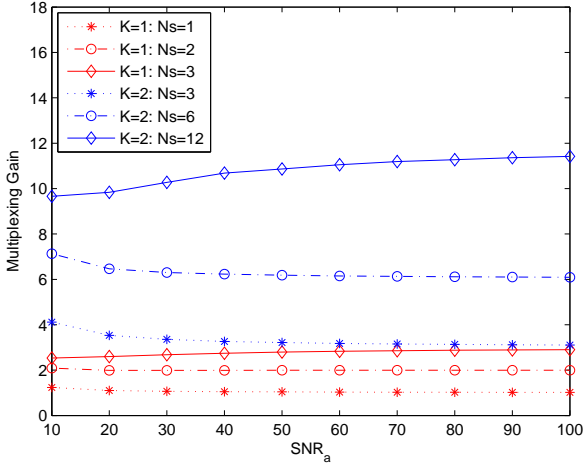}
\caption{Multiplexing gain versus $\mbox{SNR}_a$ for different numbers of data streams.}
\label{SIM4}
\end{figure}

Now we let $N_t^{({\rm rf})}=N_r^{({\rm rf})}=2N_s$\cite{Sohrabi16} and $K_r=K_t=K$, and set $L_{ij}=L=3$ and $N_r=N_t=50$. We consider the homogeneous case and define $\Psi(\bar{\gamma})=\frac{\bar{R}_{eh}(\bar{\gamma})}{\log_2\bar{\gamma}}$.  In order to verify Theorem 2, Fig.4 plots the curves of $\Psi(\bar{\gamma})$ versus $\mbox{SNR}_a=\bar{\gamma}$ for different numbers of data streams, namely, $N_s=1, 2, 3$ when $K=1$ and $N_s=3, 6, 12$ when $K=2$. It can be seen that for any given $N_s$, the function $\Psi(\bar{\gamma})$ converges to the limit value $N_s$ as $\bar{\gamma}$ grows large. This observation is expected and agrees with Theorem 2.

\subsection{Diversity-Multiplexing Tradeoff}

The previous subsection shows how much the maximal spatial multiplexing gain we can extract for a distributed mmWave-massive MIMO system while our previous work in \cite{Yue17} indicates how much the maximal spatial diversity gain we can extract. However, maximizing one type of gain will possibly result in minimizing the other. Therefore, we need to bridge between these two extremes in order to simultaneously obtain both types of gains. We firstly give the precise definition of diversity gain before we carry on the analysis.

\begin{Definition} Let $\bar{\gamma}=\frac{1}{L_s}\sum_{l=1}^{Ls}\tilde{\gamma}_l$. With an optimal power allocation, the distributed MIMO system is said to achieve spatial diversity gain $G_d$ if its average error probability satisfies

\be G_d(\bar{P}_e)=-\lim_{\bar{\gamma} \to \infty}\frac{\log_2\bar{P}_e(\bar{\gamma})}{\log_2\bar{\gamma}}.\ee

or its outage probability satisfies

\be G_d(\bar{P}_{out})=-\lim_{\bar{\gamma} \to \infty}\frac{\log_2\bar{P}_{out}(\bar{\gamma})}{\log_2\bar{\gamma}}.\ee

\end{Definition}

With the help of a result of diversity analysis in \cite{Yue17}, we can derive the following DMT result.

\begin{Theorem} Assume that both sets $\{\tilde{\ma}_r(\phi^{rl}_{ij},\theta^{rl}_{ij})\}$ and $\{\tilde{\ma}_t(\phi^{tl}_{ij},\theta^{tl}_{ij})\}$ are orthogonal vector sets when $N_r $ and $N_t$ are very large.
For a given $d \in [0, L_s]$, by using optimal power allocation, the distributed MIMO system can reach the following maximum spatial multiplexing gain at diversity gain $G_d=d$
\be \label{Gmo} G_m=\sum_{l=1}^{L_s}(1-\frac{d}{L_s-l+1})^+. \ee
\end{Theorem}

{\em Proof:} We first consider the simple case where the distributed system is the one with equal power allocation and the channel is the one with homogeneous large scale fading coefficients. The distributed system has $L_s$ available link paths in all.  For the $l$th best path, its individual maximum diversity gain is equal to $G_d^{(l)}=L_s-l+1$ \cite{Yue17}. Due to the fact that each path can not obtain a multiplexing gain of $G_m^{(l)}>1$ \cite{Tse07}, we therefore design its target data rate $R^{(l)}=r_l\log_2\bar{\gamma}$ with $0\leq r_l \leq 1$.  Then the individual outage probability is expressed as
\bee  P_{out}^{(l)}&=&\mathbb{P}(\log_2(1+\bar{\gamma}|\tilde{\beta}_l|^2)<r_l\log_2\bar{\gamma})\nnb \\
&=&\mathbb{P}(|\tilde{\beta}_l|^2<\frac{{\bar{\gamma}}^{r_l}-1}{\bar{\gamma}}). \eee
From \cite{Wang03}, \cite{Ordonez07},  the PDF of the parameter $\mu=|\tilde{\beta}_l|^2$ can be written as \be f_\mu=a\mu^{L_s-l}+o(\mu^{L_s-l})\ee where $a$ is a positive constant. So  $P_{out}^{(l)}$ can be rewritten as
\be P_{out}^{(l)}=(c\bar{\gamma})^{-(L_s-l+1)(1-r_l)}+o((\bar{\gamma})^{-(L_s-l+1)(1-r_l)}) \ee  where $c$ is a positive constant. This means that this path now can obtain diversity gain \be G_d^{(l)}=(L_s-l+1)(1-r_l).\ee Since the distributed system requires its diversity gain $G_d \geq d$, this implies that \be G_d^{(l)}=(L_s-l+1)(1-r_l) \geq d \ee or say \be \label{r_l} r_l \leq (1-\frac{d}{L_s-l+1})^+. \ee To this end, under the condition that the diversity gain satisfies  $G_d=d$, the maximum spatial multiplexing gain of the distributed system must be equal to \be \label{msmg} G_m(\bar{R}_{eh})=\sum_{l=1}^{L_s}r_l=\sum_{l=1}^{L_s}(1-\frac{d}{L_s-l+1})^+.\ee

This should be noticed that in order to achieve the maximum spatial multiplexing gain given in (\ref{msmg}), the distributed MIMO system must dynamically choose the number of data streams $N_s$ by using (\ref{r_l}).

This has proved that (\ref{Gmo}) holds under the special case. Furthermore, we readily show that for a general case, the $l$th best path can also reach a maximum diversity gain of $L_s-l+1$.  So applying (\ref{Rh}) and (\ref{Rw}) leads to the desired result. \hfill $\square$

{\em Remark 6:}  When $d$ is an integer, $G_m(d)$ can be expressed simply. In particular, $G_m(0)=L_s$ if $d=0$;  $G_m(1)=\sum_{l=1}^{L_s-1}\frac{L_s-l}{L_s-l+1}$ if $d=1$;  $G_m(L_s-1)=\frac{1}{L_s}$ if $d=L_s-1$; $G_m(L_s)=0$ if $d=L_s$. In general, if $d=L_s-N_s+1$ for a given integer $N_s \leq L_s$, then \be G_m(L_s-N_s+1)=\sum_{l=1}^{N_s-1}\frac{N_s-l}{L_s-l+1}. \ee  The function $G_m(d)$ is plotted in Fig.5. Note that $G_m(L_s-N_s)-G_m(L_s-N_s+1)=
\sum_{l=1}^{N_s}\frac{1}{L_s-l+1}.$ Generally, when $ d \in [L_s-N_s, L_s-N_s+1)$ , the multiplexing gain is given by
\be  G_m(d)=N_s- \sum_{l=1}^{N_s}\frac{d}{L_s-l+1}.  \ee

\begin{figure}[t]
\centering
\includegraphics[scale = 0.6]{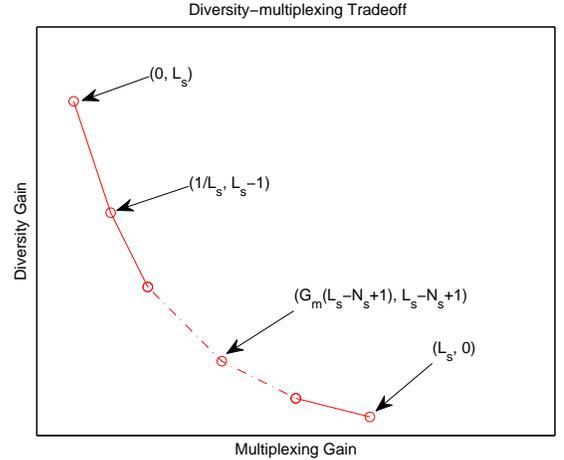}
\caption{Diversity-multiplexing tradeoff $G_m(d)$ for a general integer $d$.}
\label{DMT5}
\end{figure}

 {\em Example 1:} We set that $K_t=K_r=2$ and $L_{ij}=L=3$ . So $L_s=12$. The DMT curve with fractional multiplexing gains is shown in Fig.6.  If the multiplexing gains be limited to integers, the corresponding DMT curve is also plotted in Fig.6 for comparison.

 \begin{figure}[t]
\centering
\includegraphics[scale = 0.6]{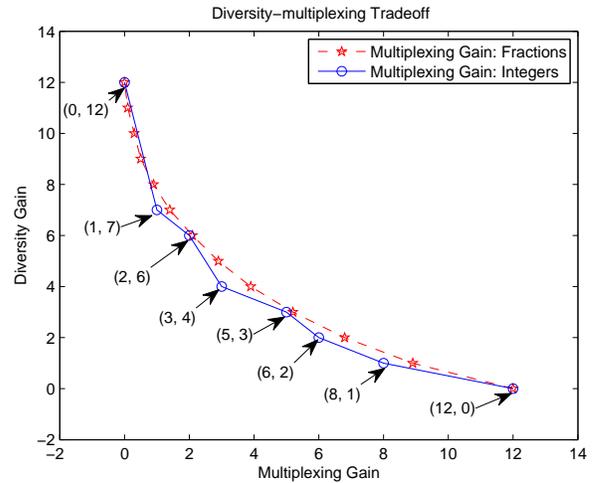}
\caption{Diversity-multiplexing tradeoff when $L=3$ and $K_t=K_r=2$.}
\label{DMT6}
\end{figure}

\section{DMT Analysis with the Conventional Partially-Connected Structure}

\begin{figure*}[t]
\centering
\includegraphics[scale=0.9]{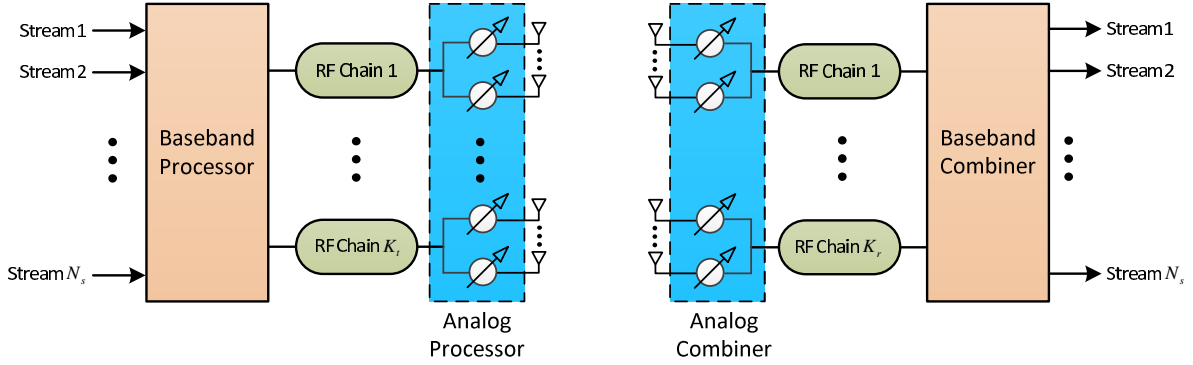}
\caption{Block diagram of a mmWave massive MIMO system with the conventional partially-connected RF architecture.}
\label{SYS7}
\end{figure*}

The previous section has analyzed the multiplexing gain for the massive MIMO system with the general fully-connected RF architecture and given a DMT characterization.
This section focuses on a massive MIMO system employing the conventional partially-connected RF architecture as illustrated in Fig. \ref{SYS7}. Here the transmitter equipped with $K_t$ RF chains sends $N_s$ data streams to the receiver equipped with $K_r$ RF chains. Each RF chain at the transmitter or receiver is connected to only one RAU. It is assumed that $N_s \leq \min\{K_t,K_r\}$. The numbers of antennas per each RAU at the transmitter and receiver are fixed as $N_t$ and $N_r$, respectively. Note that $N_t \gg N_s$ and $N_r \gg N_s$. Both the transmitter and receiver employ very small digital processors and very large analog processors, represented respectively by $\mW_t$ and $\mF_t$ for the transmitter, and $\mW_r$ and $\mF_r$ for the receiver.

As before, denote by $\ms$ the transmitted symbol vector, by $\mH$ the fading channel matrix, and by $\mn$ the noise vector. Then at the receiver the processed signal vector $\mz$ is given by (\ref{mz}), whereas $\mH$ is described as in (\ref{mH}). Due to the partially-connected RF architecture, the analog processors $\mF_t$ and $\mF_r$ are block diagonal matrices, expressed as
\be  \mF_t=\mathrm{diag}\{\mf_{t1},\mf_{t2},\ldots,\mf_{tK_t} \} \ee
and \be  \mF_r=\mathrm{diag}\{\mf_{r1},\mf_{r2},\ldots,\mf_{rK_r} \} \ee
where $\mf_{ti}$ denotes the $N_t \times 1$ steering vector of phases for the $i$th RAU at the transmitter, and $\mf_{rj}$ the $N_r \times 1$ steering vector of phases for the $j$th RAU at the transmitter.

Now let $K_m=\min\{K_r, K_t\}$ . Obviously, the distributed system at most has $K_m$ available link paths.  For the $l$th best path, its individual maximum diversity gain is denoted as $G_d^{(l)}$. In general, we can compute $G_d^{(l)}$ by an algorithm. If $L_{ij}=L$ for any $i$ and $j$, it follows from \cite{Yue17} that \be G_d^{(l)}=(K_t-l+1)(K_r-l+1)L, \; \; l=1,2,\ldots, K_m.\ee

\begin{Theorem} Consider the case that the antenna array configuration at each RAU is ULA. For a given $d \in [0, L_s]$, by using optimal power allocation, the distributed MIMO system with the partially-connected RF architecture can reach the following maximum spatial multiplexing gain at diversity gain $G_d=d$
\be \label{Gmm} G_m=\sum_{l=1}^{K_m}(1-\frac{d}{G_d^{(l)}})^+. \ee
\end{Theorem}

{\em Proof:} Following the same steps in the derivation of $G_m$ in Theorem 3, we can readily obtain (\ref{Gmm}). \hfill $\square$

{\em Remark 7:} Furthermore, we suppose that $L_{ij}=L$ for any $i$ and $j$.  Then the distributed system in the limit of large $N_t$ and  $N_r$ can reach the following maximum spatial multiplexing gain at diversity gain $G_d=d$
\be G_m=\sum_{l=1}^{K_m}(1-\frac{d}{(K_t-l+1)(K_r-l+1)L})^+. \ee

\section{DMT Analysis for the Multiuser Scenario}

\begin{figure*}[t]
\centering
\includegraphics[scale=0.8]{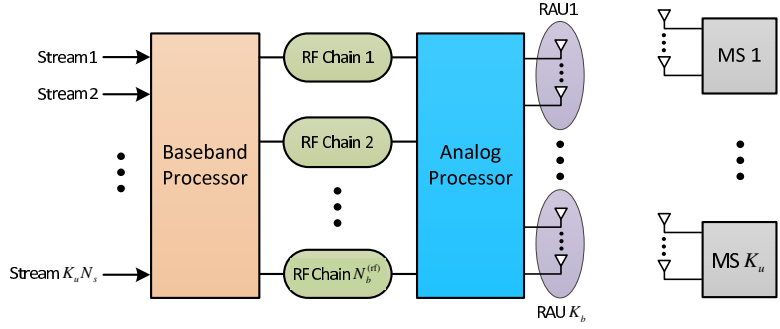}
\caption{Block diagram of a multiuser mmWave system with distributed antenna arrays.}
\label{SYS8}
\end{figure*}

This section considers the downlink communication in a multiuser massive MIMO system as illustrated in Fig. 8. Here the base station (BS) employs $K_b$ RAUs with each having $N_b$ antennas and $N_b^{({\rm rf})}$ RF chains to transmit data streams to $K_u$ mobile stations. Each mobile station (MS) is equipped with $N_u$ antennas and $N_u^{({\rm rf})}$  RF chains to support the reception of its own $N_s$ data streams. This means that there is a total of $K_uN_s$ data streams transmitted by the BS. The numbers of data streams are constrained as $K_uN_s \leq N_b^{({\rm rf})}\leq K_bN_b$ for the BS, and $N_s \leq N_u^{({\rm rf})}\leq N_u$ for each MS.

At the BS, denote by $\mF_b$ the $K_bN_b \times N_b^{({\rm rf})}$ RF precoder and by $\mW_b$ the $N_b^{({\rm rf})} \times N_sK_u$ baseband precoder. Then under the narrowband flat fading channel model, the received signal vector at the $i$th MS is given by
\be   \my_i=\mH_i \mF_b\mW_b\ms+\mn_i, \; i=1,2,\ldots, K_u \ee
where $\ms$ is the signal vector for all $K_u$ mobile stations, which satisfies $\mathbb{E}[\ms\ms^H] = \frac{P}{K_uN_s}\mI_{K_uN_s}$ and $P$ is the average total transmit power. The $N_u \times 1$ vector $\mn_i$ represents additive white Gaussian noise, whereas the $N_u \times K_bN_b$ matrix  $\mH_i$ is the channel matrix corresponding to the $i$th MS, whose entries $\mH_{ij}$ are described as in Section II. Furthermore,
the signal vector after combining can be expressed as
\be   \mz_i=\mW_{ui}^H\mF_{ui}^H\mH_i \mF_b\mW_b\ms+\mW_{ui}^H\mF_{ui}^H\mn_i, \;  i=1,2,\ldots, K_u \ee
where $F_{ui}$ is the $N_u \times N_u^{({\rm rf})}$ RF combining matrix and $\mW_{ui}$ is the $N_u^{({\rm rf})} \times N_s$  baseband combining matrix for the $i$th MS.

\begin{Theorem} Assume that all antenna array configurations for the downlink transmission are ULA. For the $i$th user, let $L_s^{(i)}=\sum_{j=1}^{K_b}L_{ij}$ and $0 \leq d^{(i)}\leq L_s^{(i)}$. In the limit of large $N_b$ and  $N_u$, the $i$th user can achieve the following maximum spatial multiplexing gain when its individual diversity gain satisfies $G_d^{(i)}=d^{(i)}$
\be G_m^{(i)}=\sum_{l=1}^{L_s^{i}}(1-\frac{d^{(i)}}{L_s^{(i)}-l+1})^+.\ee
\end{Theorem}

{\em Proof:} For the downlink transmission in a massive MIMO multiuser system, the overall equivalent multiuser basedband channel can be written as
\be \label{mHeq}
    \mH_{\rm eq}=\left[\begin{array}{llll}
  \mF_{u1}^H & \mo & \cdots & \;\;\mo \\
   \mo&\mF_{u2}^H  & \cdots & \;\;\mo \\
   \vdots & \vdots & \ddots& \;\;\vdots\\
   \mo& \mo & \cdots & \mF_{uK_u}^H
  \end{array}
   \right]\left[\begin{array}{l}
   \mH_1 \\
   \mH_2\\
    \vdots \\
  \mH_{K_u}
  \end{array}
   \right]\mF_b.
\ee On the other hand, when both $N_b$ and $N_u$ are very large, it follows easily that both BS and MS array response vector sets are orthogonal sets. Therefore the multiplexing gain for the $i$th user can depend only on the subchannel matrix $\mH_i$ and the choices of $\mF_{ui}$ and $\mF_b$. The subchannel matrix $\mH_i$ has a total of $L_s^{(i)}$ propagation paths. Similar to the proof of Theorem 2, by employing the optimal RF precoder and combiner for the $i$th user, when its diversity gain satisfies $G_d^{i}=d^{(i)}$, the user can achieve a maximum multiplexing gain of \be G_m^{(i)}=\sum_{l=1}^{L_s^{(i)}}(1-\frac{d^{(i)}}{L_s^{(i)}-l+1})^+ .\ee So we obtain the desired result. \hfill $\square$

{\em Remark 8:} Furthermore, we suppose that $L_{ij}=L$ for any $i$ and $j$ . Let $0 \leq d\leq K_bL$. Then in the limit of large $N_b$ and  $N_u$, the downlink transmission in the massive MIMO multiuser system can achieve the following maximum spatial multiplexing gain at diversity gain $G_d=d$
\be G_m=\sum_{i=1}^{K_u}G_m^{(i)}=K_u\sum_{l=1}^{K_bL}(1-\frac{d}{K_bL-l+1})^+.\ee

{\em Remark 9:} In a similar fashion, it is easy to prove that the uplink transmission in the massive MIMO multiuser system can also achieve simultaneously  a diversity gain of $G_d=d$ ($0 \leq d\leq K_bL$) and  a spatial multiplexing gain of \be G_m=K_u\sum_{l=1}^{L_s}(1-\frac{d}{K_bL-l+1})^+. \ee

\section{Conclusion}
This paper has investigated the distributed antenna subarray architecture for mmWave massive MIMO systems and provided the asymptotical multiplexing analysis when the number of antennas at each subarray goes to infinity. In particular, this paper has derived the closed-form formulas of the asymptotical available rate and spatial maximum multiplexing gain under the assumption which the subchannel matrices between transmit and receive antenna subarrays behave independently.  The spatial multiplexing gain formula shows that mmWave systems with the distributed antenna architecture can achieve potentially rather larger multiplexing gain than the ones with the conventional co-located antenna architecture. On the other hand, using the distributed antenna architecture can also achieve potentially rather higher diversity gain. For a given mmWave massive MIMO channel, both types of gains can be simultaneously obtained. This paper has finally given a simple DMT tradeoff solution, which provides insights for designing a mmWave massive MIMO system.


\end{document}